# On-Chain Archaeology of Bitcoin Oracles: Evidence of Use under Limited Observability

Giulio Caldarelli
University of Turin
ORCID-ID: 0000-0002-8922-7871
giulio.caldarelli@unito.it

## Abstract

Before Ethereum made the "oracle problem" a household term, Bitcoin already had oracles serving as feeds, key-release services, federated signers, and arbiters that carried real value on the main chain. This study traces their use and the changing evidence of oracle activity from early days through July 2026. We combine a complete census of Counterparty betting (1,149 bets), analysis of the full Bitcoin chain through block 958,628, and searches for documented keys from Reality Keys, Orisi, Bitrated, and Oraclize in an 854-million-row public-key index. We also recover DLC oracle records from an archived explorer and live Nostr relays. Two results emerge. First, early contracts remain on-chain, but many event descriptions have disappeared, and protocol encoding and API limitations complicate access to the surviving record. However, for modern DLCs, public oracle announcements can survive even when the contracts using them cannot be identified on-chain. In the script classes examined, the share of spends that reveal no script peaks at 81.9% in 2024 after excluding spends containing inscription data. Second, public registries can give a misleading picture of oracle use. In Counterparty, 95% of pre-2018 sources declaring an oracle fee were never bet on. In Bitrated, 0.1% of archived keys appear on-chain overall, compared with 10 of 19 keys captured in 2014. Sport dominates Counterparty's matched volume, while a daily price series dominates the archived DLC announcements. These findings show how protocol design and data preservation shape the historical record of Bitcoin oracle use.

## 1. Introduction

"*Oracles can potentially evaluate anything, yet the link to the block chain can always be the same*." Mike Hearn (2011b)

In 2011, Hearn proposed making Bitcoin payments conditional on events outside the network, from exchange rates to inheritance. His proposal anticipated a role for Bitcoin in agreements whose execution depended on information that the network could not establish on its own. Over the following years, developers explored this possibility through betting platforms, data feeds, key-release services, and arbitration systems. In order to pursue this research path, developing working oracles was crucial to ensure reliable real-world data collection and interoperability between platforms (Belchior *et al.*, 2022).

Research on blockchain oracles has examined the trust, security, and implementation choices involved in connecting contracts to external information (Al-Breiki *et al.*, 2020; Pasdar, Lee and Dong, 2023). For Bitcoin, historical research has reconstructed early oracle services and traced the transition from oracle co-signing to designs based on published attestations, particularly discreet log contracts (DLCs) (Caldarelli, 2023, 2026). Complementary technical work has refined DLC implementations (Le Guilly, Kohen and Kuwahara, 2022) and developed cryptographic constructions for oracle-based conditional payments (Madathil *et al.*, 2023). The scale and distribution of recorded use across these designs remain, however, unevenly documented. The existence of a service or a proposed application leaves open questions about the activity it supported, the value involved, and its concentration among providers. Addressing these questions is also complicated by differences in what each design makes public. Dryja (2017) explicitly recognized that the privacy of DLCs would make their overall use difficult to estimate from blockchain transactions. For earlier systems, identifying an application may depend on documentation and event descriptions published outside the chain. A historical comparison therefore needs to connect evidence of use with the conditions under which that evidence can be observed and interpreted.

This study addresses that gap through two research questions:

1. **RQ1.** What do the available records reveal about the scale, concentration, and applications of Bitcoin oracle use?
2. **RQ2.** How do contract design and the availability of historical sources shape what can be observed and measured?

We combine a census of Counterparty betting (1,149 submissions, 2014–2017) with analysis of the Bitcoin blockchain from genesis through July 2026 and searches for documented Reality Keys, Orisi, Bitrated, and Oraclize keys in an index of 854 million public-key occurrences. The analysis also covers DLC announcements and attestations recovered from the archived Suredbits explorer and public Nostr relays. Historical documentation and archived websites provide context for interpreting the records. Two findings stand out. Counterparty activity was concentrated in a small number of services and applications, with modest fee revenues and limited matching of offered value, even though most matched contracts reached settlement. Across the systems examined, measures of activity also depend on the role of public keys, the information disclosed by transactions, and the external records available. In the DLC collections, announcements and attestations document the information supplied by oracles, extending the public record beyond the bets and settlements visible in Counterparty. These findings add a quantitative dimension to the historical literature and show how design and source preservation affect comparisons of oracle use. For practitioners, they clarify which indicators describe the provision of oracle information and which describe its consumption. They also highlight the value of preserving event descriptions and service documentation alongside transaction records.

The remainder of the paper is organized as follows. Section 2 reviews the literature background and related work. Section 3 presents the data and methods, Section 4 reports the results, and Section 5 discusses their implications. Section 6 concludes the paper and outlines directions for further research.

## 2. Background and related work

Research on blockchain oracles has focused on how contracts obtain external information and on the trust and security trade-offs involved. Al-Breiki et al. (2020) compare mechanisms intended to make oracle services trustworthy, while Pasdar et al. (2023) review their implementation and integration with external systems. Eskandari et al. (2021) examine the stages through which oracles obtain and report external information, identifying the trust assumptions and vulnerabilities associated with different design choices. A common concern is that the guarantees of blockchain execution do not automatically extend to the accuracy of the information supplied. This problem was already relevant to Bitcoin's early development, as developers explored ways to make payments conditional on events outside the network. Caldarelli (2023) reconstructs these early efforts on Bitcoin through project documentation and developer interviews. Hearn (2011a) suggested that an oracle could authorize a transaction once an external condition was met. Orisi distributed this signing role across several participants, while Oraclize combined transaction signing with authenticated external data. Other designs made information available for contracts to use without requiring the oracle to sign each settlement. Reality Keys released the private key corresponding to an event's outcome, allowing users to construct contracts around its publication. Counterparty recorded broadcasts and betting instructions in Bitcoin transactions, with settlement governed by its own protocol.

Counterparty's betting features highlight the connection between oracle design and prediction markets. A market must, in fact, specify who can report an event's outcome and how that report determines settlement. Within this decentralized setting, Clark et al. (2014) examined how decentralized markets could organize trading and arbitration, while Sztorc (2015) proposed Truthcoin as a sidechain that combines prediction markets with incentives for honest outcome reporting. These works explored how markets could operate and resolve outcomes with less reliance on a central operator, one of the central concerns in blockchain domains. How a contract uses an oracle also affects participants' privacy. Discreet log contracts (DLCs) address this issue by allowing an oracle to publish a signature attesting to an outcome without needing to know which contracts depend on it (Dryja, 2017). The parties use the signature to settle their agreement without publishing its terms. Subsequent work develops practical implementations, including techniques for handling numerical outcomes and multiple oracles (Le Guilly, Kohen and Kuwahara, 2022). Caldarelli (2026) reviews these developments within the broader evolution of Bitcoin oracle designs. The resulting privacy also makes DLC use difficult to identify from transactions, a limitation already recognized by Dryja (2017).

This difficulty raises a broader question about what Bitcoin's transaction history can reveal about the applications using it. Meiklejohn et al. (2016) combine transaction analysis with information about known services to characterize payments and their participants. Nerurkar et al. (2021) analyze Bitcoin transactions from 2009 to 2020 to reconstruct a network of users and examine how its structure changes over time. Bartoletti and Pompianu (2017) examine OP_RETURN metadata to identify applications and measure their activity over time. Such analyses rely on connecting transaction data to knowledge of the systems that generated them. For oracle contracts, understanding a payment may also require a description of the event or condition on which it depended. Those descriptions may be published outside the blockchain, where their continued availability is less certain. Klein et al. (2014) show how online sources can disappear or change even when the publications citing them remain accessible. For Bitcoin oracles, the loss of such material can make a recorded transaction difficult to connect to the event it concerned. This creates a limit to historical reconstruction alongside the privacy built into some contract designs. Focusing exclusively on Bitcoin, the present study examines what transaction records and surviving external sources reveal about oracle use, from early services to modern DLCs.

## 3. Data and methods

This retrospective empirical study examines Bitcoin oracle activity using blockchain transactions, the Counterparty ledger, and public DLC oracle records. Table 1 summarizes these three sources and the observations they support. Historical documentation and archived websites provide information about service designs and event subjects. Bitcoin blockchain and Counterparty data were collected in July 2026, and dates are expressed in UTC.

**Table 1. The three components of the analysis and their data sources.**

| Component | Systems examined | Evidence and access | Observation units |
|---|---|---|---|
| Bitcoin blockchain | Reality Keys, Orisi, Oraclize, and Bitrated arbitration | Bitcoin Core transaction data, searched using service keys recovered from documentation and archived websites | Scripts, public-key occurrences, and transactions |
| Counterparty | Oracle feeds used by the betting protocol | Counterparty Core records of bets and on-chain broadcasts, supplemented by archived feed pages and operator websites | Bet placements, matches, settlements, and feeds |
| DLC oracle publications | Oracles represented in the Suredbits and Nostr collections | Announcements and attestations recovered from the archived Suredbits explorer and public Nostr relays | Distinct announcement payloads for Suredbits; distinct messages for Nostr |

### 3.1 The Counterparty ledger

Counterparty data were extracted in July 2026 from a local Counterparty Core node connected to a synchronized Bitcoin Core node. We retrieved valid broadcasts through the REST API and extracted bets, matches, resolutions, expirations, and block records from the node’s SQLite database. Because the ledger records successive states as additional rows, we reconstructed each bet using its first record for placement information and its latest record for final status. We similarly retained the final state of each unique matched pair. The resulting dataset contains 1,149 bets, including 1,097 protocol-valid submissions, and 397 matches. Invalid submissions were retained to distinguish rejected attempts from accepted activity. Match outcomes were checked against resolution and expiration records and the protocol event log.

Matched value was calculated as the cumulative sum of both parties' contributions across matches, counting funds again whenever they were committed to a subsequent match. Bet placements and settlements were dated using timestamps from their confirmation and resolution blocks, respectively. Each match was dated to the block confirming the second bet in the pair. User-declared deadlines were retained for comparison with event calendars; timestamps supplied within broadcasts were excluded from the activity chronology. We classified bet subjects using broadcast text, feed URLs, archived operator catalogues, and contemporary documentation. Wayback searches first targeted exact feed pages and then relevant material elsewhere on the operators' domains. Feed-specific descriptions took precedence over general platform categories; feeds without sufficient evidence were classified as unrecoverable. For xbet.io, catalogue changes were represented by grouping feeds according to their first observed activity before or after 1 September 2014. These groups distinguish the earlier soccer and World Cup offering from the later football and horse-racing offering. We also compared activity and declared deadlines with the 2014 World Cup calendar, from 12 June to 13 July. Counterparty amounts are reported in XCP, with illustrative US dollar equivalents calculated at a fixed rate of USD 2.40 per XCP. The reference prices and conversion calculations are reported in Appendix A. Figure 1 shows the summer distribution of declared deadlines.

**Figure 1.** User-declared bet deadlines by day, summer 2014: The World Cup fingerprint.

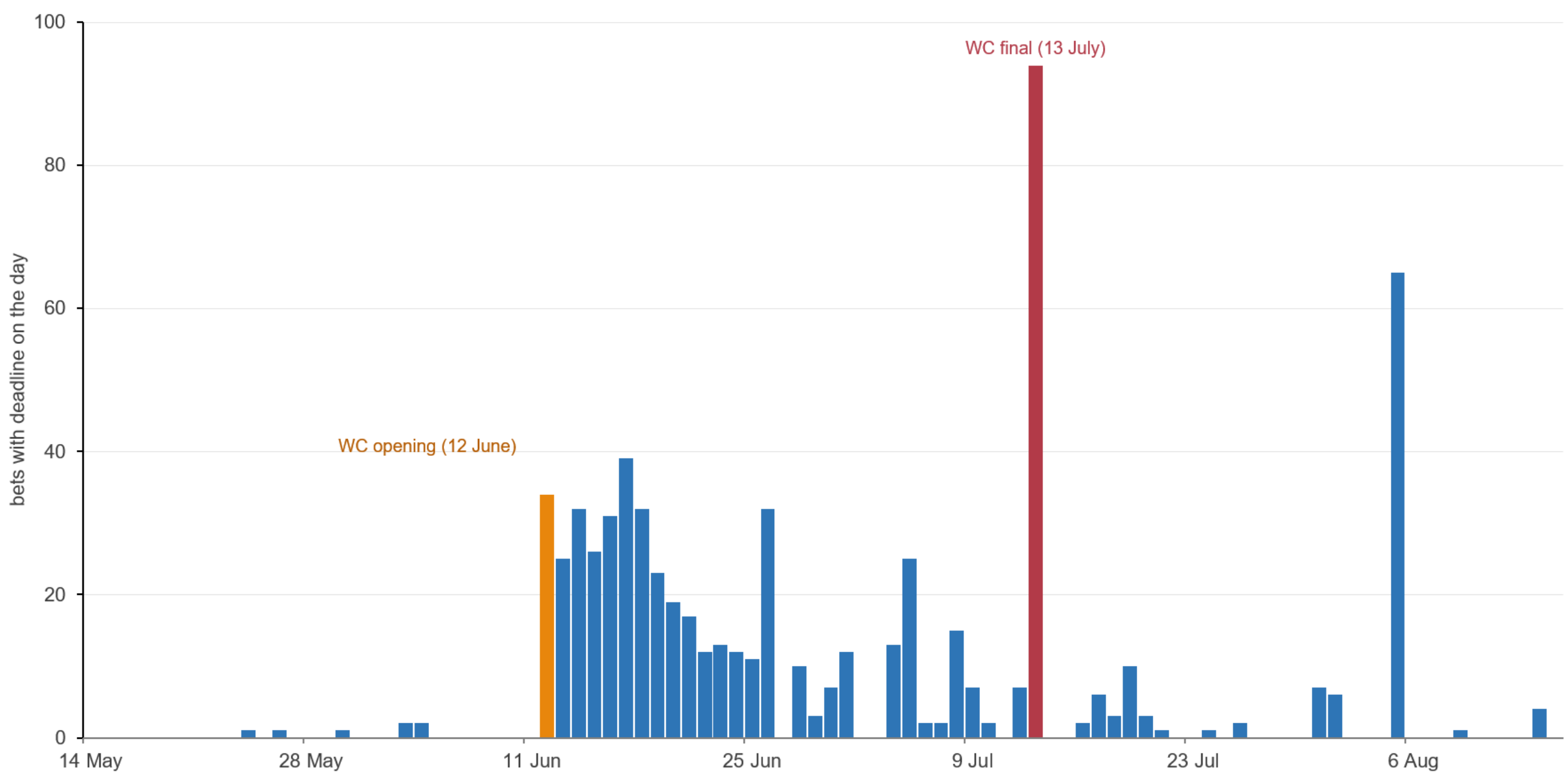

### 3.2 The Bitcoin blockchain

We extracted Bitcoin blocks 0–958,628, covering January 2009 to July 2026. Bitcoin Core supplied transactions and previous-output information, which were analyzed using DuckDB and Python. Coverage checks confirmed all 958,629 block heights, without gaps or duplicates. Annual aggregates used Bitcoin Core's mediantime timestamps. The extraction retained bare multisignature, P2SH, P2WSH, Taproot, OP_RETURN, and nonstandard outputs, together with scripts revealed by P2SH, P2WSH, and Taproot script-path spends. Nested P2SH–P2WSH spends were classified as P2WSH, while wrapped single-key spends were excluded from the revealed-script dataset. This produced 450,015,065 script-revealing spends. Taproot spending paths were also recorded.

An index of 854,077,402 public-key occurrences in multisignature inputs and outputs supported exact matching against documented service keys. Each match recorded the transaction identifier, block height, and whether the key appeared in an output or a spending input. Key occurrences and distinct transactions were counted separately. For Reality Keys, we searched the compressed forms of the six recovered public keys. Scripts were classified by their multisignature requirements, hashlocks, timelocks, and conditional branches. The HTLC-like category required a hashlock, a conditional branch, and a relative or absolute timelock. We also examined possible data storage in multisignature key fields and measured key reuse through recorded occurrences and distinct companion keys. We assessed Lightning attribution among spent 2-of-2 P2WSH outputs, including nested forms. Funding outputs were matched against historical channel announcements (Decker, 2020), while a second indicator identified commitment-transaction features described in BOLT #3 (Lightning Network Specifications, 2019). The combined count used the union of both indicators, counting overlap once. Other outputs remained unattributed. The announcement dataset ends in September 2023; the unattributed group can therefore include later or private Lightning channels alongside other applications.

Changes in script disclosure were measured using an annual opacity index defined for this study:

$$O_t = K_t / (K_t + S_t + H_t + W_t)$$

Here, $K_t$ and $S_t$ count Taproot key-path and script-path spends, while $H_t$ and $W_t$ count retained P2SH and P2WSH script reveals (single-key P2SH–P2WPKH wrappers excluded). Nested P2SH–P2WSH spends enter $W_t$. The subscript t denotes the year. Taproot key-path spends do not disclose any additional script conditions that may exist (Wuille, Nick and Towns, 2021). The index measures the share of Taproot key-path spends within the selected spending categories, including ordinary payments, and it is zero by construction before Taproot. A second version removes script-path spends containing the Ordinals envelope marker from $S_t$ to assess their influence on the measure (Rodarmor, 2023). Finally, we compared Counterparty transaction identifiers with the union of retained output and revealed-script records to measure their coverage in the blockchain extraction. Unmatched records were examined through ledger fields and a sample of transactions retrieved directly from Bitcoin Core.

### 3.3 DLC oracle publications

DLC oracle publications were examined through the archived Suredbits explorer and public Nostr relays. For Suredbits, Wayback CDX searches targeted oracle, event, announcement, and attestation pages on the main and test domains. We extracted oracle names, event descriptions, maturity dates, outcomes, and embedded payloads. Production-domain and test-domain records were kept separate, and test events were identified during classification. Deduplication by announcement payload yielded 911 distinct announcements from the production domain. We checked maturity dates against the announcement payloads and recorded whether an associated attestation payload was present in the archived pages.

Nostr collection targeted event kinds 88 and 89, used for DLC announcements and attestations in a historical, unmerged proposal (Carman, 2023). The collector was configured to query five public relays using backward pagination. Retrieved messages were deduplicated by event identifier. The resulting corpus contains 219 messages: 167 announcements and 52 attestations. These counts refer to distinct Nostr messages, which can contain repeated underlying announcements. Publisher counts use the public keys declared in announcement messages. Subjects were classified using ordered keyword rules over event names, tags, and readable payload strings, supplemented by event-specific assignments for Suredbits. Categories included prices and finance, sport, weather, and tests. Records whose subject could not be identified remained unclassified. Structural decoding was assessed separately. Category shares were calculated over announcements within each corpus. For Suredbits, the distribution was also calculated after excluding the recurring Deribit-BTC series to assess its influence. Suredbits time series used event maturity, the earliest planned attestation time, while Nostr series used the creation timestamp declared in each message.

### 3.4 Scope and interpretation

The study includes Bitrated arbitration as an application related to oracles, while services that co-sign transactions solely for wallet security and custodial gambling platforms fall outside the study's scope. OpenBazaar could not be examined through key matching because moderator keys were not recovered. Bitrated keys were obtained from archived profiles and grouped by earliest capture year. Capture dates do not establish when the keys were created. Table 2 reports the size and coverage of the datasets used in the three analyses outlined in Table 1. Public-key matching identifies documented service keys in indexed Bitcoin scripts. Counterparty records provide measures of betting activity, matched value, and settlement, while DLC records document oracle announcements and attestations. Finally, we address RQ1 through descriptive statistics on activity, value, concentration, and application domains. For RQ2, we compare how contract design and the availability of historical sources affect the identification and interpretation of oracle activity.

**Table 2. Quantitative coverage of the datasets used in the analysis.**

| Dataset | Observation unit | Coverage and sample size |
|---|---|---|
| Bitcoin blockchain extraction | Blocks, script-revealing spends, and public-key occurrences | 958,629 blocks (heights 0–958,628); 450,015,065 script-revealing spends; 854,077,402 public-key occurrences in the multisignature index |
| Documented service keys | Public keys used for exact matching against the blockchain index | Orisi: 10; Reality Keys: 6; Oraclize: 1; Bitrated: 11,261 |
| Counterparty ledger | Bet submissions and unique matched pairs | 1,149 bets, including 1,097 protocol-valid submissions; 397 matches; betting activity in 2014–2017 |
| Archived Suredbits explorer | Distinct announcement payloads | 911 announcements recovered from the production domain |
| Nostr relays | Messages deduplicated by event identifier | 219 messages: 167 announcements and 52 attestations |

## 4. Results

The results document oracle activity across Counterparty betting, transactions associated with identifiable Bitcoin services, and DLC oracle publications.

### 4.1. Counterparty Betting activity and market structure

#### 4.1.1 Activity, applications and concentration

The first bet recorded in the Counterparty ledger was placed on 2 February 2014. It was a 1 XCP wager on a feed labeled “Super Bowl Bet Feed: 1 = Seattle Won”. The activity was concentrated in 2014, which accounted for 1,105 of the 1,149 recorded bet transactions (96.2%). June alone contained 465 bets, or 40.5% of the full dataset, overlapping with the FIFA World Cup held from 12 June to 13 July. A further increase occurred in October, with 219 bets; archived xbet.io pages from this period advertised horse racing as a new feature. Only 44 bets were recorded after 2014, the last on 12 June 2017. No later submissions appear in the ledger examined through July 2026 (Figure 2).

**Figure 2.** Monthly Counterparty bet submissions

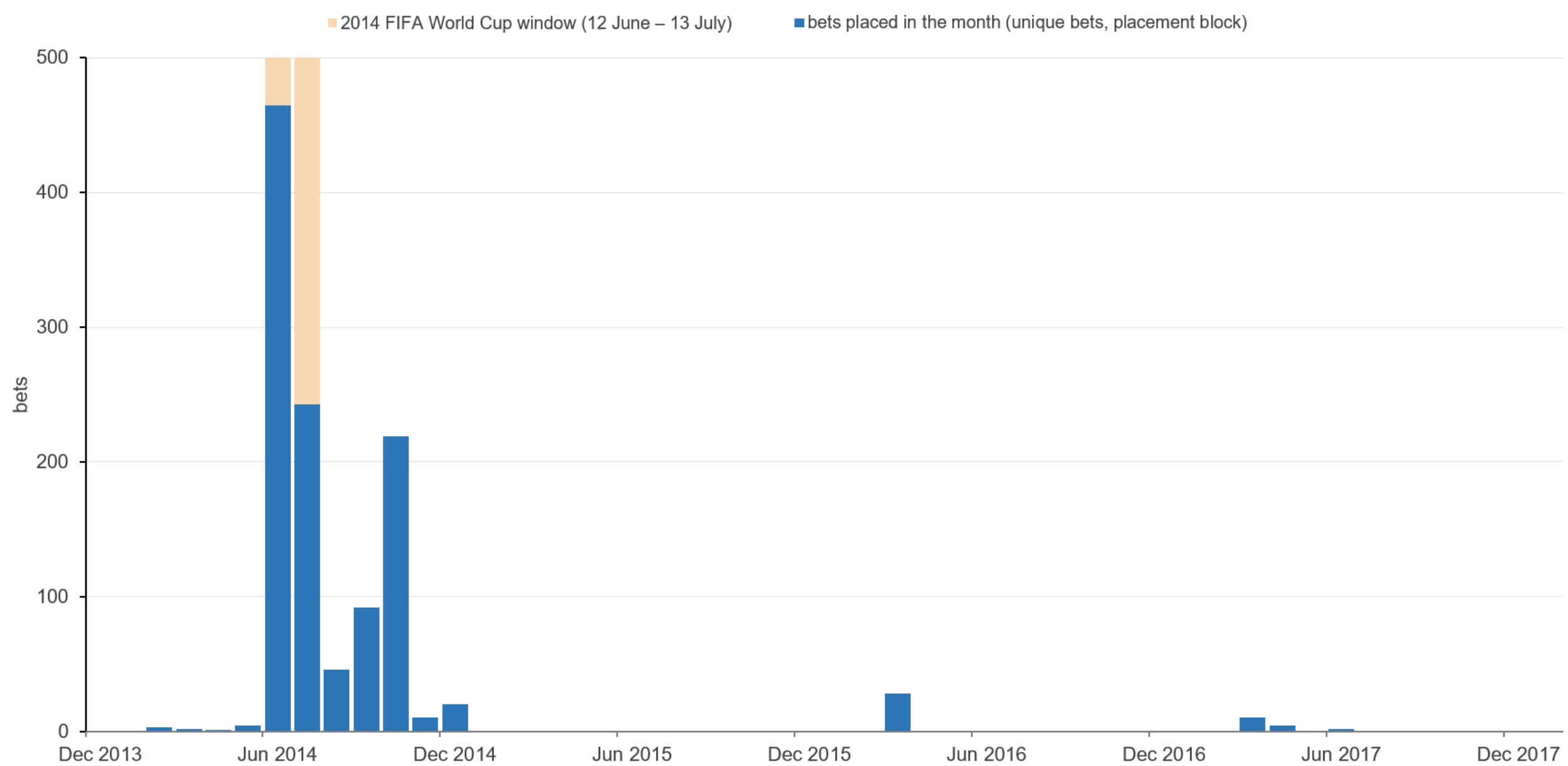

Across the observation period, sports accounted for 90.5% of matched XCP value, followed by price feeds at 8.3% and gambling feeds at 1.2% (Figure 3). Test feeds and feeds whose subjects could not be recovered together represented less than 0.1%. Most submitted bets used the binary Equal/NotEqual types, which accounted for 1,033 of the 1,149 records (89.9%). The remaining 116 were contracts for difference (CFDs), of which 86 were protocol-valid. The invalid submissions included 29 rejected with the status “invalid: CFDs temporarily disabled”.

**Figure 3.** Share of matched XCP by feed category.

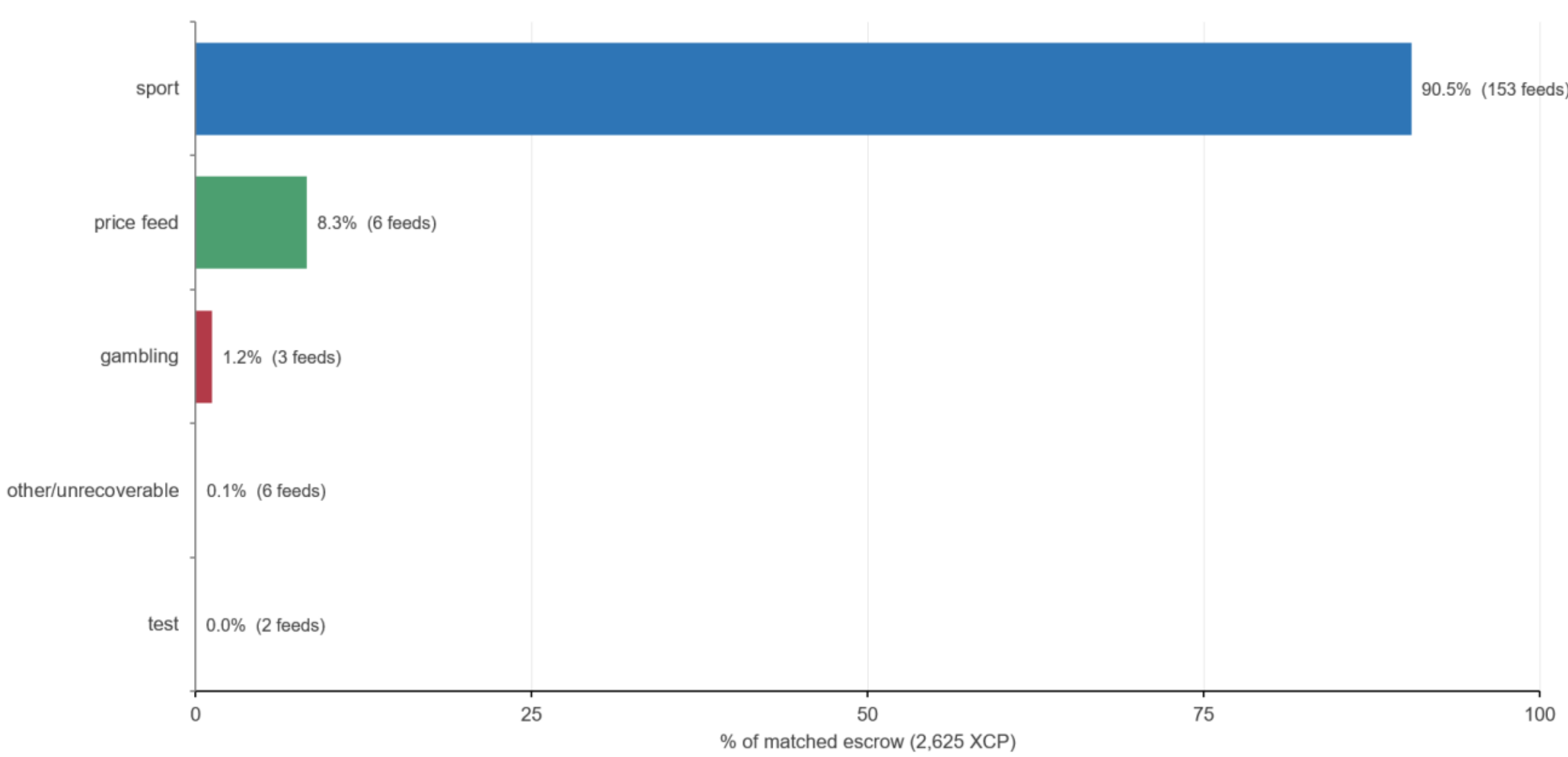

Note: Percentages refer to matched XCP value. Counts in parentheses include all 170 feeds referenced by submitted bets; 98 of these feeds had matched activity.

Activity was also concentrated among a small number of services. Across the 98 feeds with matched activity, the Herfindahl–Hirschman Index (HHI), calculated as the sum of squared percentage shares of matched value, was 711. The largest individual feed accounted for 22.5%. When feeds were grouped by operator, the HHI was 7,787 (Figure 4). The 82 matched feeds associated with xbet.io represented 87.6% of total matched value, while xcpfeeds.info accounted for 10.6%. Together, these two services accounted for 98.2% of matched value.

**Figure 4.** Matched escrow by operator (a) and Lorenz curve across the 98 matched feeds (b).

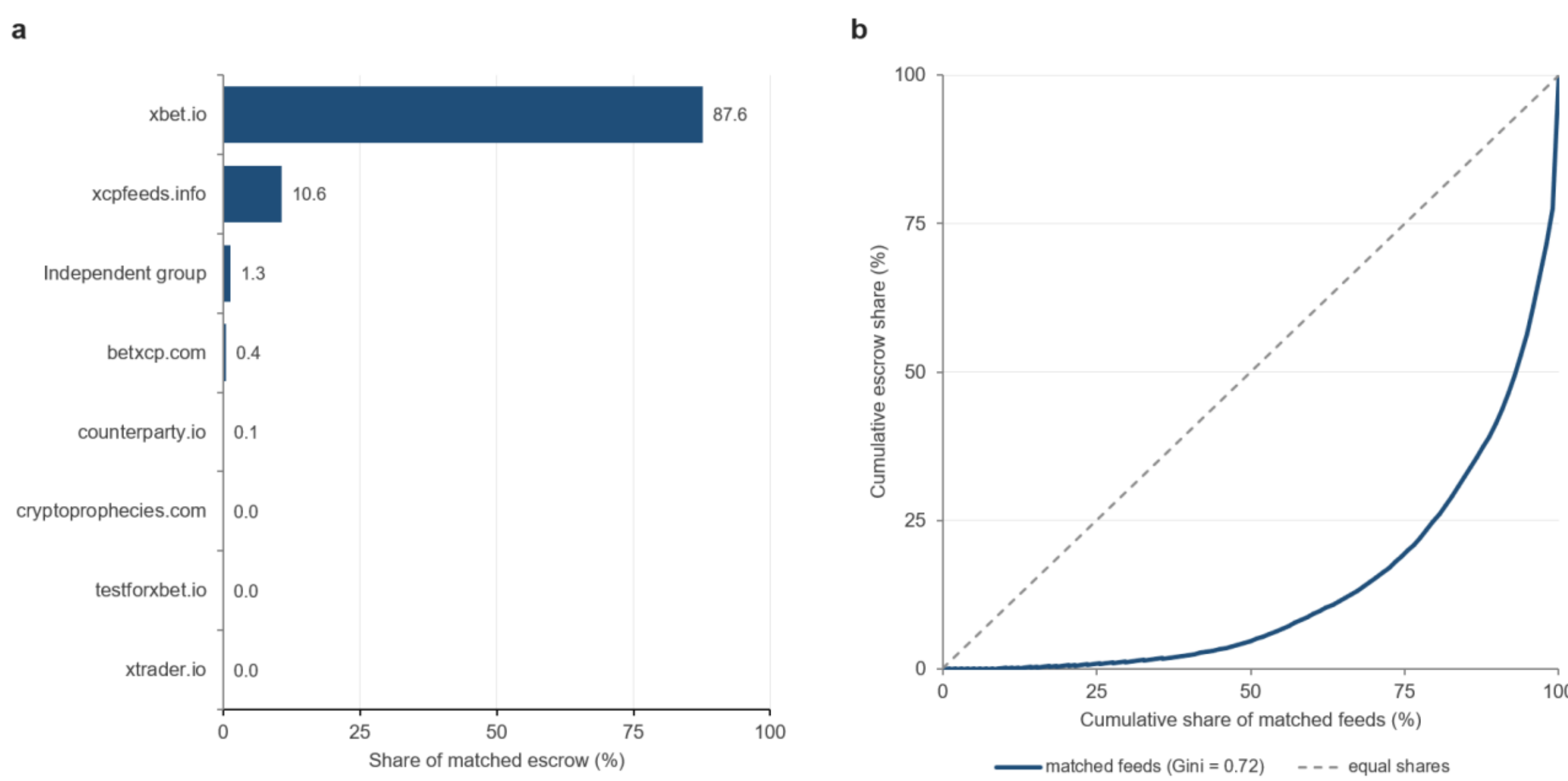

### 4.1.2 Matching, settlement and fees

The concentration of activity described above was accompanied by limited matching of submitted offers. Across the full betting history, users offered 7,158 XCP, of which 7,150 came from protocol-valid bets. Only 2,625 XCP were matched, equivalent to 36.7% of the total offered. At the illustrative exchange rate introduced in Section 3.1, these amounts correspond to approximately $17,200 offered and $6,300 matched. Of the 1,097 valid bets, 588 (54%) ended with an expired status, 38% were filled, and 8% were canceled; none remained open. Partially matched offers could also expire with an unfilled remainder. Once bets were matched, most contracts reached settlement. Of the 397 matches, 387 (97.5%) were resolved, including 70 CFD matches liquidated between 22 and 26 July 2014. Nine matches expired, returning approximately 19.8 XCP to the counterparties through the protocol's refund mechanism. One further match was dropped following a cancellation broadcast, which likewise returned the committed funds (Counterparty-Developers, 2024). Feed locking, which permanently prevents further broadcasts and new bets on a feed, was recorded for five sources (Weller, Zuber and Chiguiretor, 2019).

Fee declarations were considerably more common than betting use. Across the full broadcast record, 6,362 messages from 3,045 sources declared a non-zero fee. Among the 3,474 sources first observed before 2018, 2,965 had at least one such declaration in the recovered record, but only 153 of these (5.2%) were referenced by a bet. Across all sources, 170 feeds were referenced by bets, and 98 had matched activity (Figure 5). The limited volume of matched bets also translated into small fee revenues. Total fees across all feeds amounted to 23.7 XCP, approximately $57 at the illustrative exchange rate, while the highest-earning individual feed collected 5.9 XCP. Thus, although most matched contracts reached settlement, the recorded activity generated little revenue for feed providers.

**Figure 5.** Broadcast sources and betting activity.

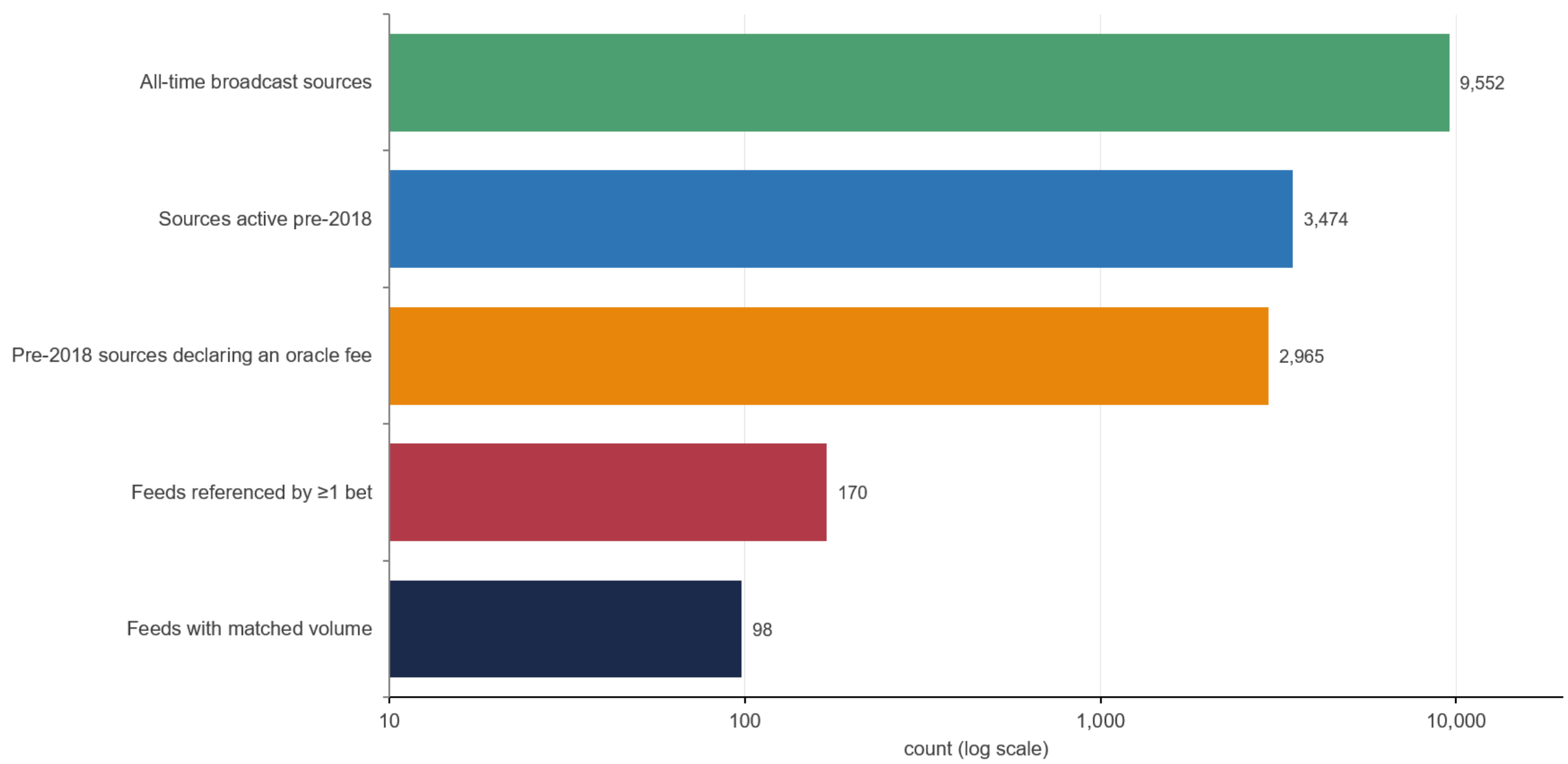

### 4.1.3 Broadcast activity and observability

Broadcasting continued after the last recorded bet in 2017. From 2022 to the July 2026 cutoff, 77,108 of the 80,066 broadcasts (96.3%) contained Broadcast Token Naming System (BTNS) commands. This protocol uses Counterparty broadcasts to issue, mint, and transfer tokens (Johnson, 2023). BTNS activity peaked at 51,220 messages in 2023 and declined to 1,283 in 2025. The same period contained 1,123 numeric feed publications from 129 sources, including a BTC–USD feed active between 2022 and 2024. A separate broadcast in June 2026 asked whether a RAREPEPE NAKAMOTO card would be sold on OpenSea before the end of the day. These publications continued without new bet transactions appearing in the Counterparty ledger examined.

Archived websites provided information about the subjects of historical bets. Counterparty broadcasts associated with xbet.io contained web addresses pointing to the service's feed pages. The July 2026 Wayback search returned no captures for 130 xbet.io feed URLs, 14 betxcp.com URLs and 11 xcpfeeds.info endpoints (Figure 6). The two control queries returned captures of the Quandl homepage and the CounterTools page on jpja.net. Broader searches recovered 174 archived xbet.io URLs, whose navigation menus documented the platform's sports catalogue and supported the category assignments in Section 4.1.1. Access to the betting records themselves also depended on the interface used. In the Counterparty Core v11.2.0 installation examined, the /v2/bets endpoint returned no records, while direct queries of the ledger database recovered 1,149 distinct bet transactions. Reconstructing this history therefore required both access to the underlying protocol records and recovery of the external descriptions needed to interpret them.

**Figure 6.** Wayback Machine search results.

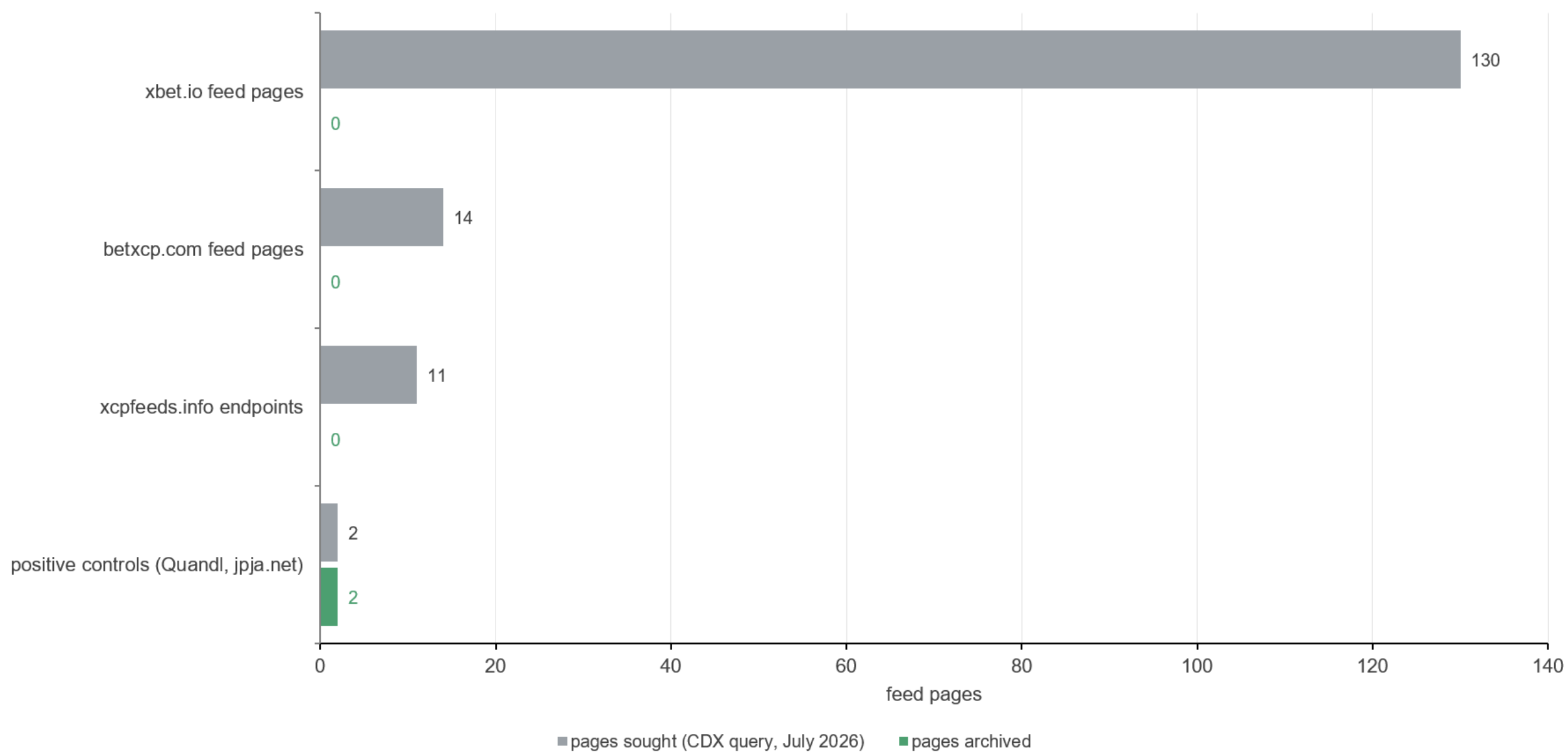

### 4.2 Bitcoin blockchain analysis

#### 4.2.1 Scripts and protocol identification

Across blocks 0–958,628, the chain analysis identified 450,015,065 spends that revealed scripts within the categories defined in Section 3.2. Of these, 316,060,091 revealed multisignature scripts. This large population provided the context for searching for oracle-related activity, although the same script structures were shared by many applications. One source of ambiguity was the use of multisignature outputs to carry data. The analysis flagged 842,241 bare-multisignature outputs containing at least one public-key value that could not be decoded as a valid curve point. Their distribution showed an early cluster in 2014–2016, a spike of 60,947 outputs in 2019, and a further increase in 2023–2026, including 216,003 outputs in 2024 and 417,638 in 2025. Reuse patterns were also difficult to interpret. The public-key index contained 11,606,109 distinct values with at least five occurrences; the three most frequent each appeared 6,375,188 times. Data-carrier placeholders also appeared in the ranking, limiting the usefulness of recurrence alone for identifying oracle services.

Counterparty illustrated how transaction formats could change within a single protocol. Transactions containing bare-multisignature outputs predominated in 2015, with 215,629 records. In 2017, 401,146 of the 401,382 ledger transactions contained an OP_RETURN output (99.94%), while bare-multisignature outputs became more common again in 2023–2024. These changes also affected recognition. A literal search for CNTRPRTY in OP_RETURN outputs identified 333 transactions, compared with 3,146,510 transactions recorded in the Counterparty ledger. Counterparty's use of ARC4 obfuscation and several output formats explains why searching for a readable protocol marker cannot recover its transaction history (Counterparty-Developers, 2026). Matching ledger transaction identifiers against the indexed Bitcoin records located 2,961,933 transactions, or 94.13%. Of the 184,577 unmatched records, 2,553 were associated with the 2014 burn period, and 178,950 fell in 2021–2025. A check of 20 transactions from the latter group found only ordinary payment output types excluded from the index. A further 3,074 unmatched transactions fell in other years and remained unclassified. The reconciliation therefore measured coverage of the selected chain records, while the sample helped explain part of the gap. Identifying transactions associated with particular oracle services required additional evidence, supplied by the documented public keys examined next.

#### 4.2.2 Transactions associated with oracle services

Matching documented public keys against the chain index located transactions associated with Orisi, Oraclize and Bitrated, while none of the recovered Reality Keys matched (Table 3). For Orisi, seven of the ten keys recovered from the repository and documentation appeared in 17 mainnet transactions between 12 June and 14 October 2014. Twelve transactions revealed 2-of-4 scripts, three revealed 2-of-3 scripts, and the remaining two revealed 2-of-2 and 1-of-1 scripts. The 17 identified inputs spent previous outputs totaling 0.02037468 BTC, with a median value of 0.001 BTC, equivalent to approximately $12 in total using the historical price anchors in Appendix A. The three keys from repository test fixtures produced no matches.

Oraclize's documented service key, recovered from an archived copy of contracts_create.js, appeared in 106 distinct transactions dated 2015–2016, all containing 2-of-3 scripts. These transactions document the inclusion of the service key in mainnet scripts. The founder reported that the Bitcoin-side activity was conducted for testing purposes (Caldarelli, 2023). For Bitrated, 11 of the 11,261 public keys recovered from archived profiles appeared in 56 distinct transactions, all with 2-of-3 scripts and dated 2013–2015. The overall match rate of 0.10% concealed substantial differences between archive cohorts (Figure 7). Ten of the 19 keys first captured in 2014 matched (52.6%), as did 11 of the 97 captured by the end of 2015 (11.3%). All matched keys belonged to these early cohorts. Bitrated's security documentation, archived on 17 March 2015, distinguishes profile identity keys from the separately derived keys used in contracts (Bitrated, no date). The reported percentages describe direct matches among profile keys grouped by their earliest archived capture. The six compressed public keys recovered from three archived Reality Keys example facts produced no matches in the multisignature index. The client supported scripts with explicit oracle keys and a construction combining an oracle key with a participant's key (Edgar and Delrey, 2014). In the latter construction, the script contains the combined public key.

**Table 3. Direct matching of documented service keys against the Bitcoin public-key index.**

| Service | Keys searched | Keys matched | Distinct transactions | Observed script types | Transaction years |
|---|---|---|---|---|---|
| Orisi | 10 | 7 | 17 | 2-of-4, 2-of-3, 2-of-2, 1-of-1 | 2014 |
| Oraclize | 1 | 1 | 106 | 2-of-3 | 2015–2016 |
| Bitrated | 11,261 | 11 | 56 | 2-of-3 | 2013–2015 |
| Reality Keys | 6 | 0 | 0 | No matches found | Not applicable |

**Figure 7.** Bitrated profile keys by year of earliest recovered archive capture.

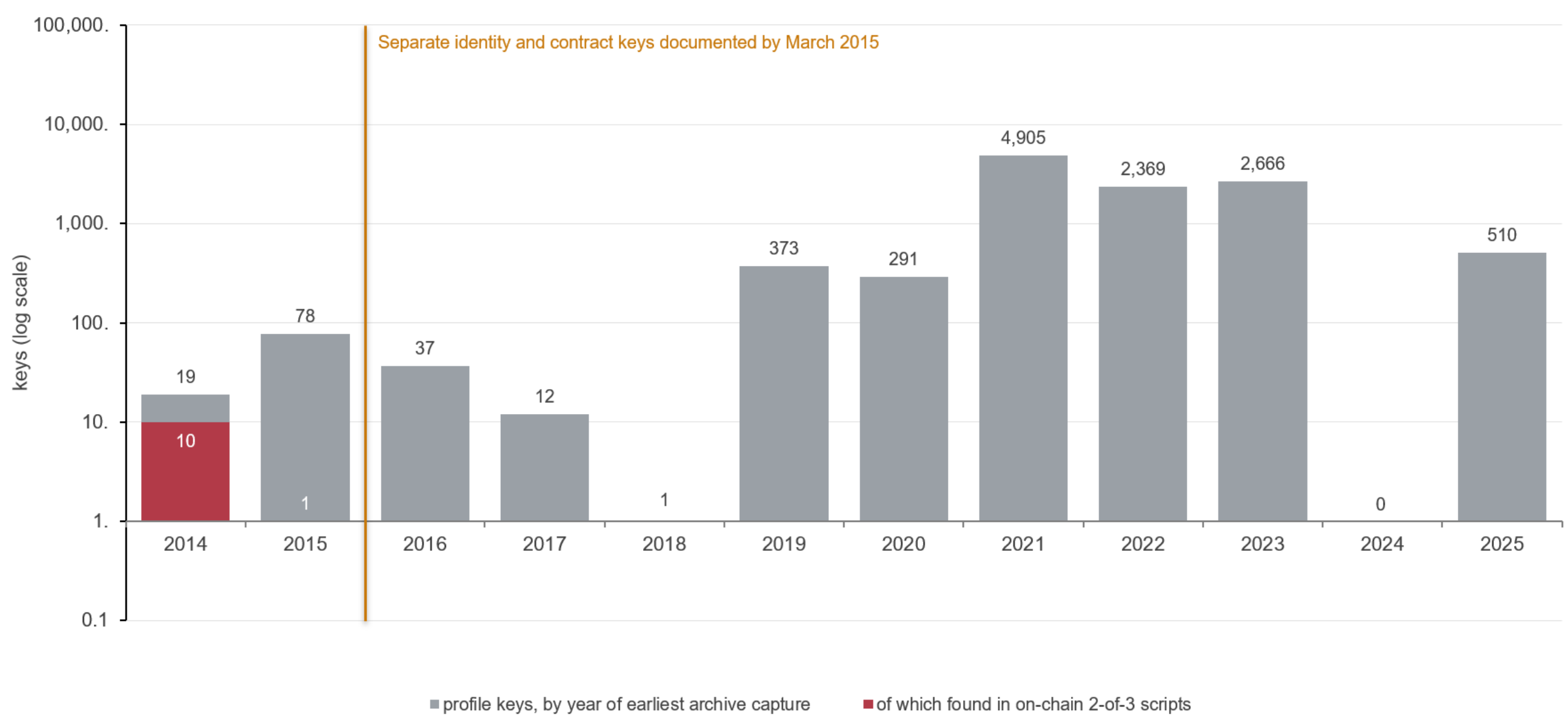

### 4.2.3 Script disclosure and limits of attribution

The spending path also affected the information available for attribution. Taproot key-path spends disclose a signature without revealing a spending script (Wuille, Nick and Towns, 2021). Within the categories defined in Section 3.2, their share rose from 2.6% in 2022 to 67.4% in 2024, before declining to 52.8% in 2025 and 50.8% in the partial 2026 record (Table 4; Figure 8). The series was sensitive to the treatment of inscription-related scripts. Between 2023 and the 2026 cutoff, 92–98% of Taproot script-path spends contained the Ordinals marker. Excluding these spends from the denominator raised the key-path share to 48.6% in 2023, 81.9% in 2024, 76.2% in 2025, and 70.1% in 2026. The difference was particularly large in 2023, when the share including these spends was 23.3%. Both series describe script disclosure within the selected population, which also includes ordinary payments.

**Table 4. Annual spend counts and Taproot key-path shares.**

| Year | Key-path | Script-path | With ord marker | P2SH | P2WSH | All selected spends (%) | Excluding ord (%) |
|---|---|---|---|---|---|---|---|
| 2021 | 21,850 | 18 | 0 | 4,468,166 | 34,288,155 | 0.06 | 0.06 |
| 2022 | 961,802 | 1,018 | 3 | 2,748,078 | 33,327,211 | 2.60 | 2.60 |
| 2023 | 23,679,017 | 54,180,135 | 52,946,571 | 2,955,859 | 20,896,973 | 23.28 | 48.56 |
| 2024 | 83,785,768 | 23,878,971 | 21,987,554 | 3,096,057 | 13,547,059 | 67.40 | 81.89 |
| 2025 | 57,378,874 | 34,480,592 | 33,323,812 | 2,057,381 | 14,740,564 | 52.81 | 76.17 |
| 2026 (partial) | 18,543,242 | 10,922,353 | 10,064,720 | 615,196 | 6,446,829 | 50.76 | 70.07 |

**Figure 8.** Taproot key-path share among selected spends, 2014–2026.

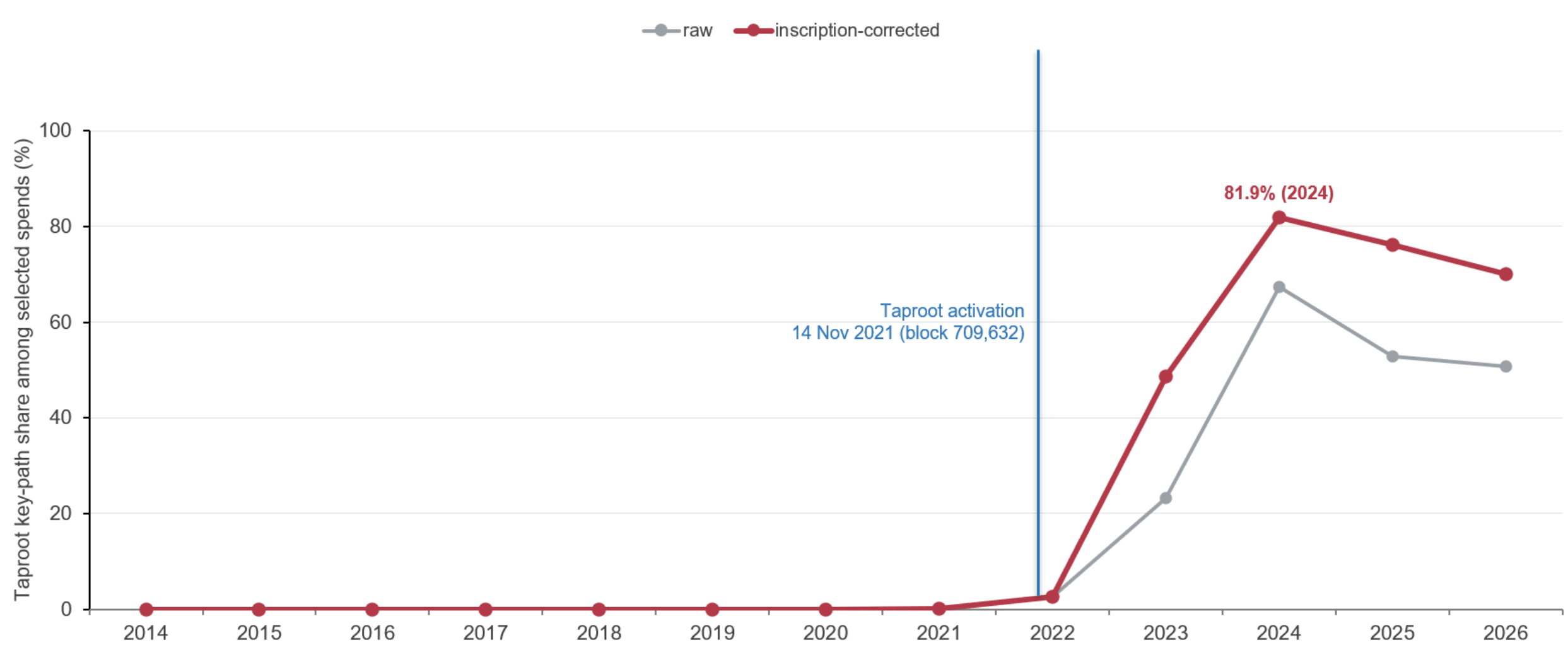

Among the 45,860,478 spent 2-of-2 P2WSH outputs, 482,165 matched publicly announced Lightning channels. Commitment patterns identified a further 290,104 spends outside that group. Together, the two indicators identified 772,269 spends (1.68%), including 197,739 identified by both methods (Table 5; Figure 9). The remaining 45,088,209 spends (98.32%) formed the unattributed group. This group accounted for at least 96.05% of the annual totals.

**Table 5. Lightning attribution among spent 2-of-2 P2WSH outputs.**

| Year | Spent outputs | Announcement matches | Commitment-only matches | Both methods | Unattributed | Unattributed (%) |
|---|---|---|---|---|---|---|
| 2017 | 1,875,699 | 0 | 9 | 0 | 1,875,690 | 100.00 |
| 2018 | 4,629,078 | 353 | 34,890 | 331 | 4,593,835 | 99.24 |
| 2019 | 8,908,054 | 59,036 | 22,681 | 29,465 | 8,826,337 | 99.08 |
| 2020 | 7,386,362 | 44,257 | 12,510 | 17,368 | 7,329,595 | 99.23 |
| 2021 | 6,830,019 | 90,952 | 12,278 | 27,264 | 6,726,789 | 98.49 |
| 2022 | 6,559,342 | 146,730 | 16,434 | 57,133 | 6,396,178 | 97.51 |
| 2023 | 4,337,145 | 98,927 | 31,570 | 45,282 | 4,206,648 | 96.99 |
| 2024 | 2,322,177 | 29,137 | 62,486 | 13,884 | 2,230,554 | 96.05 |
| 2025 | 2,013,984 | 10,964 | 63,752 | 6,139 | 1,939,268 | 96.29 |
| 2026 (partial) | 998,618 | 1,809 | 33,494 | 873 | 963,315 | 96.46 |
| **Total** | **45,860,478** | **482,165** | **290,104** | **197,739** | **45,088,209** | **98.32** |

**Figure 9.** Lightning attribution among spent 2-of-2 P2WSH outputs.

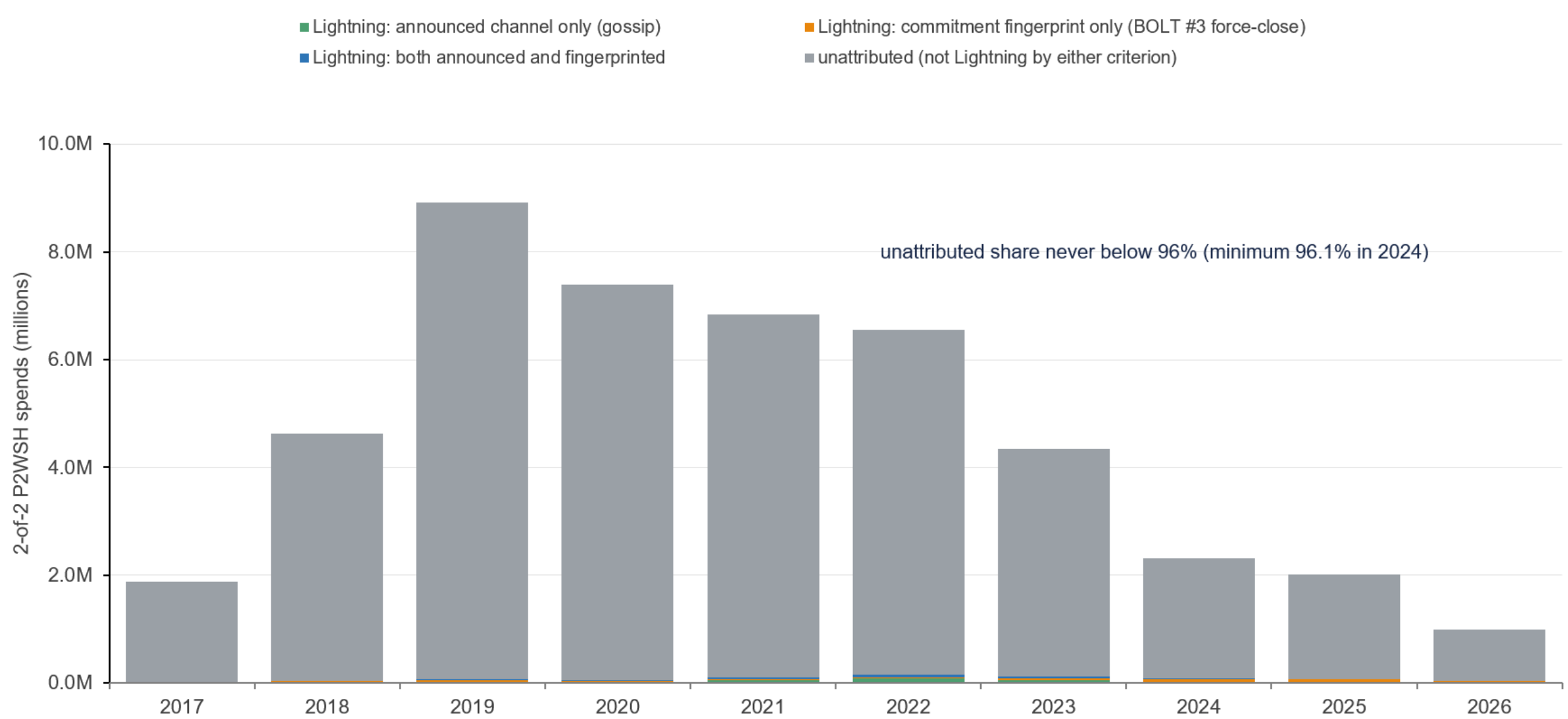

### 4.3 DLC oracle publications

#### 4.3.1 The Suredbits explorer archive

The analysis covered 3,270 records extracted from archived announcement and event pages of the Suredbits explorer, with capture dates between 2019 and 2023. The production-domain records yielded 911 distinct announcement payloads. Of these, 673 (73.9%) were associated with attestation records in the recovered pages. The recorded event maturities were concentrated in 2021 (282 announcements) and 2022 (446), followed by 2023 (174). Announcements appeared under several publisher names. The most frequent was suredbits-oracle-bot, with 564 announcements, followed by Pythia (77) and Suredbits (43). Other names included @Oracle_Pythia_ (30), Zees Sees (19), Chris_Stewart_5 (17) and benthecarman (14).

Within this record, price and finance accounted for 649 announcements (71.2%), while sport accounted for 170 (18.7%), mainly football fixtures. The remaining announcements concerned current affairs and community questions, test events, randomness, Bitcoin network events, politics, and weather. Much of the concentration in price and finance came from the recurring Deribit-BTC reference series, which contributed 597 announcements, or 65.5% of the corpus. Excluding this series left 314 announcements, with sport accounting for 54.1% (170) and price and finance for 16.6% (52), reversing their ranking (Figure 10). The publication mix was therefore strongly influenced by one recurring series.

**Figure 10.** Composition of the recovered Suredbits announcements.

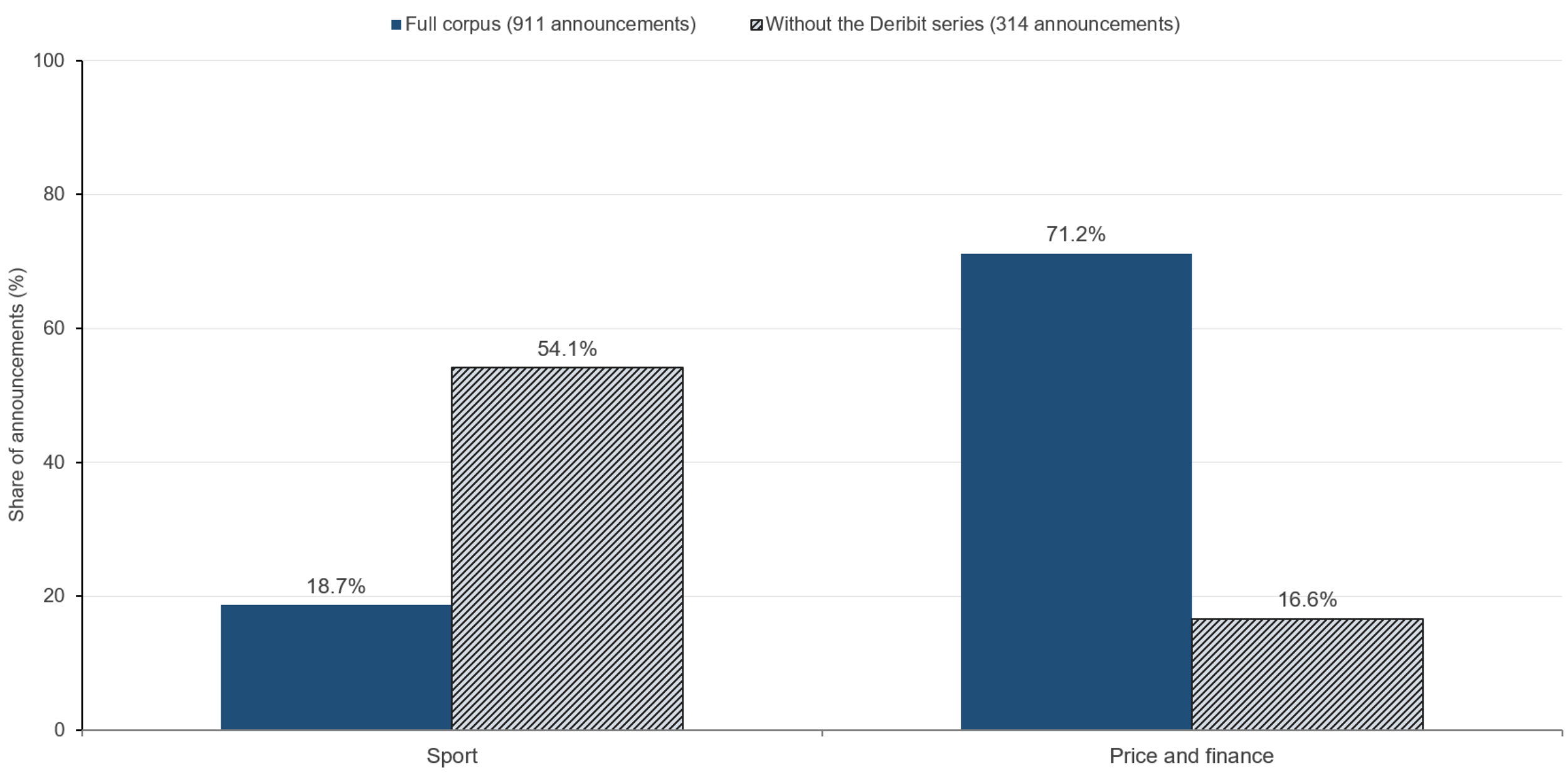

### 4.3.2 Nostr relay publications

The relay collection returned 219 distinct Nostr messages, comprising 167 announcement messages and 52 attestation messages using kinds 88 and 89 of the historical DLC proposal (Carman, 2023). The announcements were published under 21 distinct public keys. Recorded creation dates fell in 2024 (14 messages) and 2026 (205), with none in 2025. Most announcements were classified as tests or demonstrations, accounting for 95 of the 167 records (56.9%). A further 33 (19.8%), all under one publishing key, described delivery or return outcomes. The same key also appeared in 12 attestation messages with delivery-related labels.

Other announcements concerned Bitcoin prices (13 records across six publishing keys), current affairs (9), weather (3), sport (2), politics (2) and Bitcoin network conditions (1). Six contained generic winner labels, while three remained unclassified under the keyword rules. The decoder parsed 165 of the 167 announcement payloads and 49 of the 52 attestation payloads. These payloads contained the DLC message bodies directly. In all 214 decoded messages, the embedded oracle key matched the public key recorded in the Nostr author field.

#### 4.3.3 Comparison across sources

Table 6 brings together the topics identified across the three collections. Sports account for 83.5% of Counterparty bets with identified feeds and 90.5% of matched XCP value. Price and finance represent 71.2% of Suredbits announcements, largely due to the recurring Deribit-BTC series. Excluding this series makes sport the largest category, accounting for 54.1% of the remaining announcements. In the Nostr collection, tests and demonstrations predominate, followed by delivery and return labels published under a single key.

Counterparty contributes both bet counts and matched value to the comparison, while the DLC columns describe announcement counts. The 179 transactions associated with Orisi, Bitrated and Oraclize are reported separately in Table 3, which summarizes the script matches for documented service keys. Figure 11 presents annual record counts dated by block inclusion for Counterparty, event maturity for Suredbits, and message creation for Nostr. The accompanying Taproot series shows the key-path share among the selected Bitcoin spends.

**Table 6. Topic distributions in the recovered collections (%).**

| Topic | Counterparty bets | Counterparty matched XCP | Suredbits announcements | Nostr announcement messages |
|---|---:|---:|---:|---:|
| Sport | 83.5 | 90.5 | 18.7 | 1.2 |
| Price and finance | 11.7 | 8.3 | 71.2 | 7.8 |
| Gambling and randomness | 1.7 | 1.2 | 1.9 | 0 |
| Delivery and return | 0 | 0 | 0 | 19.8 |
| Politics and regulation | 0 | 0 | 1.0 | 1.2 |
| Weather | 0 | 0 | 0.5 | 1.8 |
| Bitcoin network events | 0 | 0 | 1.3 | 0.6 |
| Current affairs and community | 0 | 0 | 3.1 | 5.4 |
| Test and demonstration | 2.4 | <0.1 | 2.3 | 56.9 |
| Other or unclassified | 0.8 | 0.1 | 0 | 5.4 |

Note: Counterparty bet percentages are calculated over the 1,145 submissions with identified feeds.

**Figure 11.** Annual oracle records by source and Taproot key-path share among selected spends.

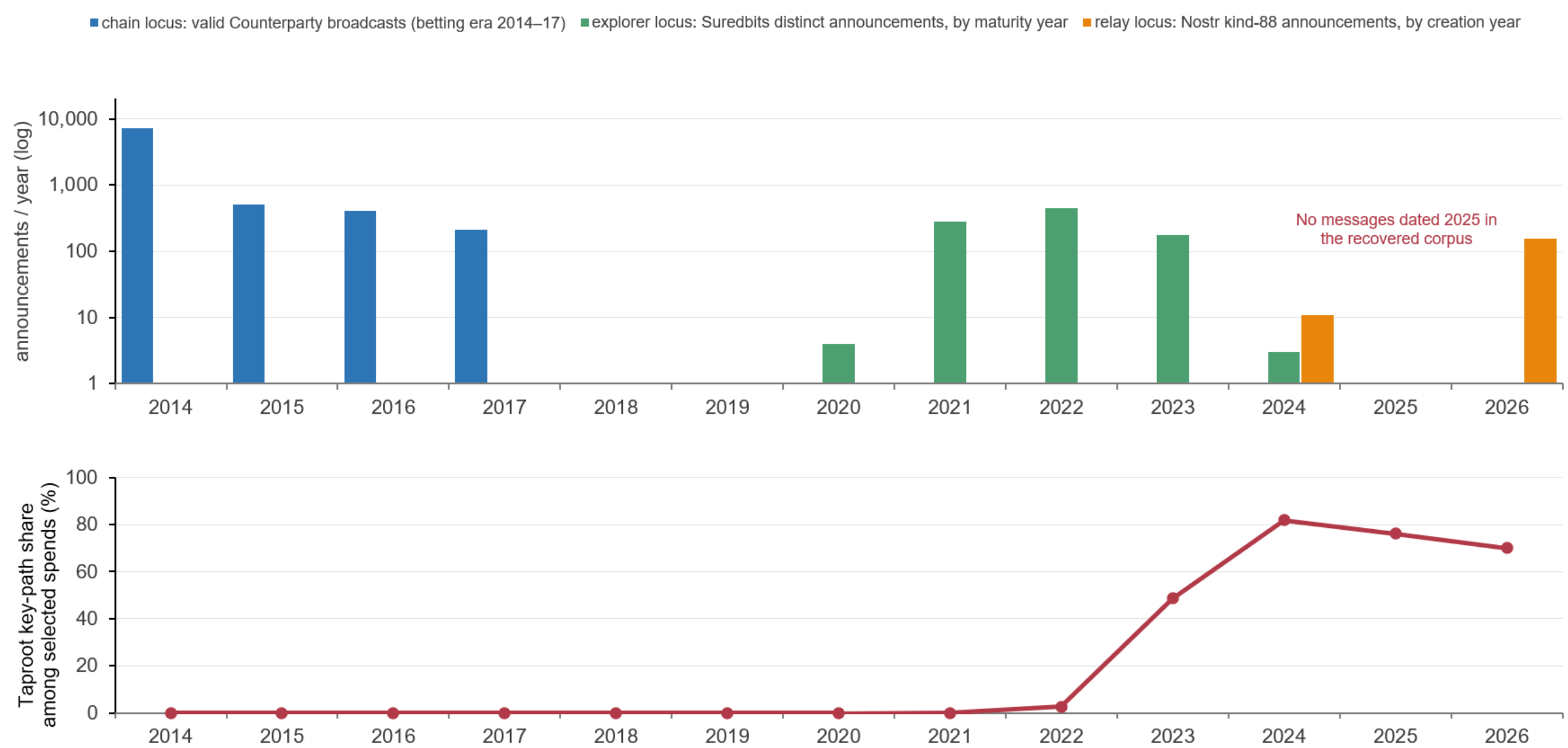

## 5. Discussion

### 5.1 Patterns of oracle use

The Counterparty findings illustrate the distinction between matching and settlement made by Clark et al. (2014) in their design of decentralized prediction markets. Only 36.7% of offered value was matched, while 97.5% of matches reached settlement. Most broadcast sources were never referenced by a bet. Together, these observations show why feed availability, participation in contracts, and settlement completion require separate assessment. Clark et al. also identify expected future fees as an incentive for reliable arbitration. Counterparty provides an empirical perspective on this incentive mechanism. The modest fees recorded show that the realized financial reward for oracle provision remained small even though most matched contracts reached settlement.

Concentration provides a further connection with empirical research on oracle adoption. Kaleem and Shi, (2021) found that a small number of projects accounted for most Chainlink price-feed consumption on Ethereum. In Counterparty, concentration becomes apparent when individual feeds are grouped by service, with two services accounting for 98.2% of matched value. The concentration occurs at different levels in the two studies, involving consuming applications on Ethereum and services organizing bets and feeds on Counterparty. Both findings underline the importance of identifying the services behind protocol activity when assessing the diversity of use. The recurrence of sport and financial prices across Counterparty and Suredbits also connects these records to established betting and market-data applications (Table 6). The financial component is consistent with the prominence of price feeds documented by Kaleem and Shi (2021). Our findings add a historical perspective within Bitcoin, where executed CFDs document early financial use and later DLC announcements show continued provision of financial data. The tests and delivery-related announcements found on Nostr broaden the range of published topics, while placing much of that diversification within experimentation.

The share of publications devoted to a topic depends partly on how frequently its data are updated. For example, a Chainlink documentation page (Chainlink, no date) explains that price feeds publish updates when a specified time interval has elapsed or when the price has changed beyond a predefined threshold. More frequent updates therefore do not necessarily mean that more contracts are using the data. A comparable issue arises when interpreting the Suredbits archive, where the recurring Deribit price series strongly shapes the topic distribution. Sport becomes the largest category when that series is excluded. Comparisons of application domains therefore need to account for publication schedules and the contribution of recurring series. For researchers and service providers, this means assessing the breadth of applications alongside the concentration and frequency of the activity supporting them.

### 5.2 Interpreting the historical record

Interpreting oracle activity requires evidence about the services and purposes behind recorded transactions. Meiklejohn et al. (2016) linked Bitcoin activity to known services through direct interactions and publicly disclosed identifiers. Bartoletti and Pompianu (2017) likewise used web documentation to identify protocols and their application domains. Our findings show how this dependence on external information extends to historical research. Recovered keys connect transactions to particular oracle implementations, while documentation helps establish the roles those keys played. The history of a service therefore matters when interpreting the activity associated with its identifiers.

This relationship becomes especially clear when transactions refer to external web pages. Klein et al. (2014) show how disappearing or changing web resources can prevent later readers from recovering the context of a reference. The xbet.io case illustrates a related problem within a blockchain application. Its broadcasts preserved web addresses pointing to feed pages, while archived menus helped reconstruct the service’s sports catalogue. Blockchain records and archived websites thus contribute different information. The blockchain records protocol activity, while the website helps explain how the service presented and organized its applications. Developers’ accounts provide another connection between transactions and their historical setting. Caldarelli (2023) reconstructs early Bitcoin oracles through interviews, forums and technical documentation, emphasizing the circumstances of their development. Our findings complement that account with dated evidence of implementation. The Oraclize transactions are consistent with the testing described by its founder and document the period and volume of the activity identified. For Orisi, the reported difficulties in relaying and mining non-standard transactions help place the observed transactions within the conditions developers faced. Reading these sources together connects the measured activity to the purposes and operating conditions described in earlier research.

These comparisons also give practical relevance to the preservation of service documentation. For oracle developers and archivists, useful records include the public keys used in each period, protocol versions, event descriptions, and copies of service interfaces. Preserving these materials alongside transaction and publication records maintains the connections needed to interpret them later. Historical reconstruction therefore contributes to an understanding of what recorded activity represented within a particular implementation and period.

### 5.3 Contract design and the measurement of use

Contract design affects both the information published on-chain and its usefulness for identifying applications. Grötschla et al. (2026) reconstruct Lightning channel activity by combining public channel announcements, Bitcoin transactions and additional heuristics. Their findings show that cooperative and unilateral closures leave different transaction patterns. Our 2-of-2 analysis highlights the complementary problem that the same script structure can serve several applications (Table 5). Together, these findings support combining script analysis with records from the relevant protocol when measuring application use. A shared transaction structure alone provides insufficient grounds for assigning the remaining, unclassified activity to oracle contracts. Taproot changes the information available for this attribution. Under BIP 341, key-path spending does not publish a script or disclose whether a script could also have been used to spend the output (Wuille, Nick and Towns, 2021). Our results trace the use of this spending option within the selected categories, with its share peaking at 81.9% in 2024 after excluding identified inscription-related spends (Table 4). This documents a change in script disclosure across a population that also includes ordinary payments. Comparing the series with and without inscriptions shows how the composition of blockchain activity affects aggregate disclosure statistics. Historical comparisons therefore need to account for changes in both spending design and the types of activity being counted.

For DLCs, the implications extend to the relationship between an oracle and its users. Dryja (2017) anticipated that estimating total DLC use would be difficult because the proposed transactions resembled Lightning transactions. His design also allows an oracle to publish information without knowing which counterparties use it. The Suredbits and Nostr collections provide an empirical way to study the oracle's activity under this separation. Their records support analysis of the topics, timing and concentration of announcements, alongside the outcomes reported through attestations. This complements transaction-based measurement by documenting the information services available to contract participants. For researchers and analytics providers, comparisons across protocols should therefore specify whether they count oracle publications, identifiable contract transactions or committed value, and explain how each system's design connects these measures.

**6. Conclusion**

This study reconstructs Bitcoin oracle activity across early services and more recent DLC publications. The Counterparty census documents 1,149 bet submissions between 2014 and 2017, with activity concentrated in 2014 and matched value dominated by two services. Most matched contracts reached settlement, although much of the offered value remained unmatched and fee revenues were modest. Key matching identified 179 transactions associated with Orisi, Oraclize, and Bitrated, while the Suredbits and Nostr collections extend the record to later oracle publications. Sport and financial prices recur across these records, alongside tests and experimentation with other applications. The findings show how contract design shapes the evidence available for measuring use. Revealed scripts and documented keys can connect transactions to particular services, while shared script structures and spending paths that disclose no script affect attribution. DLC announcements and attestations provide a complementary record of the information published by oracles. Interpreting this evidence requires distinguishing publication activity from the contracts that consume it and considering how the relationship between the two changes across implementations.

Future research could extend this historical comparison to additional oracle services and publication channels. Further archival work could also provide more detailed event descriptions, supporting finer comparisons of the applications served by oracles. For DLCs, a particularly useful direction would be to link public announcements and attestations to contract records voluntarily shared by service providers and participants. This would allow researchers to examine which published events are used in funded contracts, how much value is committed, and how activity is distributed across applications. Comparing these documented uses across implementations and periods could clarify how oracle activity evolves alongside changes in contract design. Such research would strengthen our understanding of how oracle technologies support economic activity as privacy protections change the evidence available to study their use.

## Data availability

The processed datasets supporting this study are provided in the Supplementary Material released upon journal publication. They cover Counterparty betting and broadcast activity, the Bitcoin script and service-key analyses, and DLC oracle publications recovered from the archived Suredbits explorer and Nostr relays.

## References

Al-Breiki, H. *et al.* (2020) 'Trustworthy Blockchain Oracles: Review, Comparison, and Open Research Challenges', *IEEE Access*, 8, pp. 85675–85685. Available at: https://doi.org/10.1109/ACCESS.2020.2992698.

Bartoletti, M. and Pompianu, L. (2017) 'An Analysis of Bitcoin OP_RETURN Metadata', in *Lecture Notes in Computer Science (including subseries Lecture Notes in Artificial Intelligence and Lecture Notes in Bioinformatics)*. Springer Verlag, pp. 218–230. Available at: https://doi.org/10.1007/978-3-319-70278-0_14.

Belchior, R. *et al.* (2022) 'A Survey on Blockchain Interoperability: Past, Present, and Future Trends', *ACM Computing Surveys*, 54(8), pp. 1–41. Available at: https://doi.org/10.1145/3471140.

Bitrated (no date) *Bitrated security (identity key and hardened HD master key)*, *bitrated.com*. Available at: https://www.bitrated.com/security (Accessed: 9 March 2026).

Caldarelli, G. (2023) 'Before Ethereum. The Origin and Evolution of Blockchain Oracles', *IEEE Access*, 11, pp. 50899–50917. Available at: https://doi.org/10.1109/ACCESS.2023.3279106.

Caldarelli, G. (2026) 'From Multi-sig to DLCs: Modern Oracle Designs on Bitcoin'. Available at: http://arxiv.org/abs/2602.09822.

Carman, B. (2023) *NIP-88: Discreet log contracts over Nostr. Draft proposal, GitHub pull request no. 919*. Available at: https://github.com/nostr-protocol/nips/pull/919 (Accessed: 19 March 2025).

Chainlink (no date) *Decentralized Data Model*, *docs.chain.link*. Available at: https://docs.chain.link/architecture-overview/architecture-decentralized-model (Accessed: 24 July 2026).

Clark, J. *et al.* (2014) 'On Decentralizing Prediction Markets and Order Books', *13th Annual Workshop on the Economics of Information Security*, pp. 1–21. Available at: http://weis2014.econinfosec.org/papers.php.

CoinMarketCap (2014) *Historical snapshots, 2 February – 19 October 2014 (weekly)*, *CoinMarketCap.com*. Available at: https://coinmarketcap.com/historical/20140713/ (Accessed: 5 January 2026).

Counterparty-Developers (2024) *Counterparty-core*, *Github*. Available at: https://github.com/CounterpartyXCP/counterparty-core/blob/v11.2.0/counterparty-core/counterpartycore/lib/messages/bet.py (Accessed: 13 January 2026).

Counterparty-Developers (2026) *Counterparty Protocol Specification*, *docs.counterparty.io*. Available at: https://docs.counterparty.io/docs/advanced/protocol/ (Accessed: 21 March 2026).

Decker, C. (2020) 'Lightning Network gossip datasets (2020)'. Zenodo. Available at:

https://doi.org/10.5281/zenodo.4088530.

Dryja, T. (2017) *Discreet Log Contracts*, *MIT Media Lab*. Available at: https://static1.squarespace.com/static/6675a0d5fc9e317c60db9b37/t/66e4597b7f23866561a64a95/1726241147904/discreet+log+contracts+paper.pdf (Accessed: 18 October 2025).

Edgar, E. and Delrey, C. (2014) *Realitykeysdemo.py*, *Github.com*. Available at: https://github.com/edmundedgar/realitykeys-examples/blob/master/realitykeysdemo.py (Accessed: 6 January 2023).

Eskandari, S. *et al.* (2021) 'SoK: oracles from the ground truth to market manipulation', in *Proceedings of the 3rd ACM Conference on Advances in Financial Technologies*. New York, NY, USA: ACM, pp. 127–141. Available at: https://doi.org/10.1145/3479722.3480994.

Grötschla, F. *et al.* (2026) 'On the Lifecycle of a Lightning Network Payment Channel', in *Financial Cryptography and Data Security*. Springer, pp. 1–16. Available at: https://doi.org/10.1007/978-3-032-00492-5_1.

Le Guilly, T., Kohen, N. and Kuwahara, I. (2022) 'Bitcoin Oracle Contracts: Discreet Log Contracts in Practice', *IEEE International Conference on Blockchain and Cryptocurrency, ICBC 2022*, pp. 1–8. Available at: https://doi.org/10.1109/ICBC54727.2022.9805512.

Hearn, M. (2011a) *Contracts*, *BitcoinWiki*. Available at: https://en.bitcoin.it/w/index.php?title=Contract&oldid=13637 (Accessed: 2 December 2022).

Hearn, M. (2011b) *Oracle Transactions*, *Bitcointalk.org*. Available at: https://bitcointalk.org/index.php?topic=31635.0 (Accessed: 12 April 2022).

Johnson, J. (2023) *CIP28 - Broadcast Token Name System (BTNS)*, *forums.counterparty.io*. Available at: https://forums.counterparty.io/t/cip28-broadcast-token-name-system-btns/6597 (Accessed: 28 June 2025).

Kaleem, M. and Shi, W. (2021) 'Demystifying Pythia: A Survey of ChainLink Oracles Usage on Ethereum', in *Lecture Notes in Computer Science (including subseries Lecture Notes in Artificial Intelligence and Lecture Notes in Bioinformatics)*. Springer Science and Business Media Deutschland GmbH, pp. 115–123. Available at: https://doi.org/10.1007/978-3-662-63958-0_10.

Klein, M. *et al.* (2014) 'Scholarly Context Not Found: One in Five Articles Suffers from Reference Rot', *PLoS ONE*. Edited by J. Bar-Ilan, 9(12), p. e115253. Available at: https://doi.org/10.1371/journal.pone.0115253.

Lightning Network Specifications (2019) *BOLT: Basis of Lightning Technology*, *Github*. Available at: https://github.com/lightning/bolts (Accessed: 15 June 2023).

Madathil, V. *et al.* (2023) 'Cryptographic Oracle-Based Conditional Payments', *30th Annual Network and Distributed System Security Symposium, NDSS 2023*, (March), pp. 1–18. Available at: https://doi.org/10.14722/ndss.2023.24024.

Meiklejohn, S. *et al.* (2016) 'A fistful of Bitcoins', *Communications of the ACM*, 59(4), pp. 86–93.

Available at: https://doi.org/10.1145/2896384.

Nerurkar, P. *et al.* (2021) 'Dissecting bitcoin blockchain: Empirical analysis of bitcoin network (2009–2020)', *Journal of Network and Computer Applications*, 177, p. 102940. Available at: https://doi.org/10.1016/j.jnca.2020.102940.

Pasdar, A., Lee, Y.C. and Dong, Z. (2023) 'Connect API with Blockchain: A Survey on Blockchain Oracle Implementation', *ACM Computing Surveys*, 55(10), pp. 1–39. Available at: https://doi.org/10.1145/3567582.

Rodarmor, C. (2023) *Ordinal theory handbook: Inscriptions*, *docs.ordinals.com*. Available at: https://docs.ordinals.com/inscriptions.html (Accessed: 19 December 2024).

Sztorc, P. (2015) *Truthcoin Peer-to-Peer Oracle System and Prediction Marketplace*, *bitcoinhivemind.com*. Available at: https://bitcoinhivemind.com/papers/truthcoin-whitepaper.pdf (Accessed: 15 February 2023).

Weller, D., Zuber, I. and Chiguiretor (2019) *Protocol Specification | Counterparty*, *Github.com*. Available at: https://github.com/CounterpartyXCP/Documentation/blob/master/Developers/protocol_specification.md (Accessed: 25 January 2023).

Wuille, P., Nick, J. and Towns, A. (2021) *BIP 341: Taproot: SegWit version 1 spending rules*, *GitHub*. Available at: https://github.com/bitcoin/bips/blob/master/bip-0341.mediawiki (Accessed: 23 December 2022).

*Primary sources recovered archivally (Internet Archive, Wayback Machine): Reality Keys (realitykeys.com, 2013–2018, fact API endpoints); Bitrated (bitrated.com, arbiter profile pages); Oraclize (oraclize.it, 2015–2016, contracts_create.js); xbet.io, betxcp.com, xcpfeeds.info (2014 navigation pages); oracle.suredbits.com (2019–2023 explorer snapshots); Counterparty founders' and operators' BitcoinTalk threads (2014).*

**Appendix A. Currency conversion references and calculations**

US dollar equivalents provide an indication of the economic scale of the recorded activity. XCP and BTC amounts remain the primary monetary measures.

For Counterparty, a fixed illustrative rate of USD 2.40 per XCP is applied to the aggregates for the full observation period. This rate is close to the price of USD 2.3646 recorded on 13 July 2014 (CoinMarketCap, 2014). Table A1 shows the resulting equivalents, rounded as reported in the main text.

**Table A1.** Counterparty amounts converted at USD 2.40 per XCP.

| Measure | Amount (XCP) | Approximate equivalent (USD) |
|---|---:|---:|
| Offered value | 7,158 | 17,200 |
| Matched value | 2,625.225 | 6,300 |
| Oracle fees | 23.7 | 57 |

For Orisi, the previous outputs spent by the 17 identified inputs total 0.02037468 BTC. Grouped by spending date, 0.019 BTC relates to June–July 2014 and 0.00137468 BTC to October 2014. These amounts are converted using illustrative reference prices of USD 626.50 and USD 389.55 per BTC, respectively, from the CoinMarketCap snapshots of 13 July 2014 and 19 October 2014. The calculation gives USD 12.44, reported as approximately USD 12 in the main text.